# On the derivation of the magnetocaloric properties in ferrimagnetic spinel $Mn_3O_4$

Subhash Thota, Francois Guillou, Vincent Hardy, Alexandre Wahl and Wilfrid Prellier*

Laboratoire CRISMAT, CNRS UMR 6508, ENSICAEN, 6 Boulevard du Maréchal Juin,

F-14050 Caen Cedex, France

Jitendra Kumar

Materials Science Programme, Indian Institute of Technology Kanpur, Kanpur-208016, India

## Abstract

Large magnetocaloric effect has been observed in $Mn_3O_4$ around its ferrimagnetic transition at $T_N = 42.75$ K. Field-induced isothermal entropy changes ($\Delta S$) were derived from both magnetic and calorimetric techniques. The maximum $|\Delta S|$ and adiabatic temperature change ($\Delta T_{ad}$) at $T_N$ are 11 J $kg^{-1}$ $K^{-1}$ and 1.9 K, respectively, for a magnetic field change of 20 kOe. Moreover, it is found that the complex magnetic phase transitions taking place below $T_N$ produce additional —but smaller— features on $\Delta S(T)$.

PACS number(s): 75.50.Gg, 75.25.+z, 75.30.Kz, 75.60.Ej

*Corresponding author. Email: wilfrid.prellier@ensicaen.fr

# I. Introduction:

Research activity in the area of magneto-thermodynamics (MT) has received rapid impetus in the recent past, mainly because of the discovery of giant magnetic entropy changes in the $Gd_5(Si_xGe_{1-x})_4$ compounds.[1, 2] Many efforts are underway to discover materials showing large magneto-caloric-effect (MCE) under moderate applied magnetic fields[3], so that magnetic cooling technology may become a reality in the near future.[4,5] MCE is an important phenomena of MT which manifests itself as an isothermal magnetic entropy change ($\Delta S$) or an adiabatic temperature change ($\Delta T_{ad}$) when the magnetic material is exposed to a varying magnetic field.[6] Refrigeration based on the MCE is advantageous in that it is an environmentally friendly and energy efficient alternative to the commonly used vapor-cycle refrigeration.[6] It was found that giant MCE is exhibited by various intermetallic materials such as $Gd_5(Si,Ge)_4$ [1,2], $Mn(As,Sb)$ [7], $La(Fe,Si)_{13}$ [8,9], $MnFe(P,As)$[10], Heusler alloys [11,12] and some other compounds (see reviews Ref. 3, 13-15 and references therein). In the case of transition metal oxides, most of the MCE studies so far were focused on mixed-valency rare-earth based manganese oxides[3,16], since some of them exhibit large $|\Delta S|$ values close to room temperature which are comparable with Gd.[17]

Up to now, no studies were devoted to the magnetocaloric properties of binary manganese oxides (like $MnO_2$, $Mn_2O_3$, $Mn_3O_4$ and $Mn_5O_8$). This family of compounds combines several features favorable for applications; they are cheap, harmless and have a high chemical stability.[18] Also, these oxides are potential candidates for i) rechargeable lithium batteries, ii) catalysts, and iii) soft magnetic materials for transformers cores.[18] Among all these antiferromagnetic oxides, the spinel $Mn_3O_4$ is the only one which exhibits a long-range ferrimagnetic ordering[19], i.e. a transition at which significant MCE can be expected. It belongs to the class of normal-spinel ($AB_2O_4$) structured compounds, in which the tetrahedral A sites are occupied by a divalent cation ($Mn^{2+}$) while all the octahedral B sites are occupied by trivalent cations ($Mn^{3+}$). At 1445 K, this compound undergoes a tetragonal to cubic phase transition (c/a ~1.16) resulting from a collective Jahn-Teller distortion within the $e_g$ orbitals of $Mn^{3+}$ ($t_{2g}^3 e_g^1$).[20] Below 42 K~$T_N$, a complex succession of rearrangements takes place in the magnetic structure. Despite the longstanding history[19-24], this issue is still the subject of intense research activity.[25-27] At the present time, the most widely accepted picture of its magnetic ordering is [25-27]: (i) At $T_N$, $Mn_3O_4$ exhibits a transition from the paramagnetic state towards a non-collinear ferrimagnetism of Yafet-Kittel type.[28] That means a triangular structure in which the $Mn^{2+}$ spins order ferromagnetically along [110] (cubic settings) while the $Mn^{3+}$ are symmetrically canted around [-1 -1 0 ], leading to a net moment antiparallel to that of $Mn^{2+}$; (ii) At $T_1$~39 K this coplanar spin structure lying in (1 -1 0) evolves to a spiral-type ordering related to a conical distribution of the $Mn^{3+}$ spins around [110] (this type of magnetic structure is incommensurate with the chemical lattice); (iii) Finally, at $T_2$~33 K, the spin order recovers a planar type structure. However, there is a doubling of the magnetic cell with respect to the chemical one, induced by a subdivision of the $Mn^{3+}$ into two sets. While the $Mn^{2+}$ spins still order ferromagnetically along [110], the $Mn^{3+}$ spins exhibit a canting angle of $\approx$ 69° with respect to [-1 -1 0]. Very recently, Kim *et al.* reported that this magnetic transition is accompanied by a structural transition from tetragonal to monoclinic symmetry, which is highly sensitive to the strength and orientation of the external magnetic field.[27]

In this context, our motivation to investigate the magnetocaloric properties of $Mn_3O_4$ is twofold. First, from a fundamental point of view, this compound offers an opportunity to study the MCE associated with a wide variety of magnetic or magnetostructural transitions. Second, from an application point of view, the existence of such a sequence of transitions could be expected to generate MCE over a wide temperature range. This aspect is a favorable condition to obtain large refrigeration capacity. Furthermore, our investigations also provide a comparative study on the reliability of the calorimetric and magnetic methods which are commonly used to derive the isothermal entropy changes.

## II. Experimental details:

Oxalate based sol-gel process has been chosen to synthesize $Mn_3O_4$ polycrystalline sample. The initial product is manganese oxalate dihydrate ($MnC_2O_4\cdot2H_2O$), prepared using manganese acetate tetrahydrate [$Mn(CH_3COO)_2\cdot4H_2O$] and oxalic acid [$C_2H_2O_4\cdot2H_2O$] as precursors with ethanol as a solvent. Thermal decomposition of $MnC_2O_4\cdot2H_2O$ at 500°C for 2 h duration yields $Mn_2O_3$ as a major product; further annealing of this sample at 1100°C for 4 h in air leads to the formation of $Mn_3O_4$ powder. This powder is grounded in an agate mortar, pressed into bars and finally sintered at 1100°C for 12h. The crystallinity and structure has been studied using a Panalytical X'Pert Pro diffractometer using Cu Kα radiation, which indicate monophasic nature of this compound consistent with the hausmannite $Mn_3O_4$ reported in the literature. A detailed study of the formation mechanism has been reported elsewhere.[18, 29] A superconducting quantum interference device (SQUID) based magnetometer (Quantum Design, MPMS-XL5) has been used for the temperature dependent magnetization measurements $M_H(T)$ in low-field, while magnetic isotherms $M_T(H)$ up to 90 kOe were recorded by means of an extraction technique in a "physical property measurement system" (PPMS, Quantum design). The heat-capacity curves $C(T)$ (in H=0 and H=20 kOe) were recorded from 2 to 100 K by using a semi-adiabatic relaxation technique in the PPMS. To stress the comparison between the calorimetric and magnetic methods, the paper is focused on the data corresponding to a field change from zero to

20 kOe. Note also that H = 20 kOe is the typical value to be considered for potential applications using permanent magnets.

## III. Results and discussion:

The location of the above mentioned transitions ($T_N$,$T_1$,$T_2$) on the magnetization and heat capacity data are reported in the insets of Figure 1 and 2a, respectively. Apart from $T_N$ which has a clear signature in both cases, the features associated to $T_1$ and $T_2$ are faint. To reveal them clearly, we display the temperature dependence of the first derivative of the magnetization curve (–dM/dT) recorded upon cooling in a small magnetic field of 200 Oe (inset of Figure 1) while inset of Figure 2a shows the temperature dependence of C/T in zero field. There is a good consistency between these two results, both leading to $T_N \approx$ 42.75 K, $T_1 \approx$ 39.75 K and $T_2 \approx$ 34.25 K. The positions of these transitions are in line with those previously reported in the literature.[23,25-27]

In order to derive the magnetic entropy change, we first used the common method based on the Maxwell equation applied to a series of isothermal magnetization curves. These isothermal magnetization curves were recorded as follows: (i) Each of these curves was preceded by a zero-field-cooling, ensuring to start from a virgin magnetic state; (ii) The temperature spacing between two successive $M_T$(H) curves was quite large out of the region of the transitions (δT=5K), and it was decreased to a small value inside it (δT=2K); (iii)  The data were recorded upon increasing the magnetic field from 0 to a value of $H_{Max}$. Two series were recorded, with $H_{Max}$ = 20 kOe and $H_{Max}$ = 90 kOe. The results of ΔS(T) for magnetic field changes equal or lower than 20 kOe were checked to be the same in both cases. Making use of the Maxwell equation with such data, the temperature dependence of ΔS associated to a magnetic field change from 0 to H is usually derived from the formula:[6]

$$\Delta S(T,H) = \int\limits_0^H \left[ \frac{\partial M(T,H')}{\partial T} \right]_{H'} dH' \qquad (1a)$$

It must be emphasized that this relationship assumes that M(T,H) can be regarded as a state function, i.e. taking a single value independent of the magnetothermal history. Accordingly, the Maxwell equation can be safely used around a second-order transition, whereas it is in principle inapplicable to a first-order transition.[30-42] In the latter case, the analysis of the magnetic data that is required to estimate ΔS is much more complex. In the recent years, various models have been proposed to address this issue which is still intensively debated.[31-33,35,37,38,40,42]

In the present case of $Mn_3O_4$, the MCE primarily takes place around $T_N$ which is a second-order transition. We thus derived ΔS(T) from the Maxwell equation, by using a slightly modified form of this equation which is more tractable in practice:[15,41,43,44]

$$\Delta S(T,H) = \frac{\partial}{\partial T} \int\limits_0^H M(T,H')dH' \qquad (1b).$$

The main panel of Figure 1 displays the -ΔS(T) curve obtained at H = 20 kOe, when considering the $M_T(H)$ curves recorded upon field increasing. As expected for a ferrimagnetic transition, one observes a large peak centered around $T_N$, which reaches a maximum value of ~ 7.5 $Jkg^{-1}K^{-1}$. The overall shape of -ΔS(T) was found to be the same for large values of H, the main difference being the height of the peak which reaches ~ 13 $Jkg^{-1}K^{-1}$ and ~ 18 $Jkg^{-1}K^{-1}$ for H = 50 and 90 kOe, respectively. Beside this main peak, the second prominent feature of -ΔS(T) in Figure 1 is the presence of a crossing from positive to negative values as temperature is decreased below 20 K. Similar features were reported in various types of manganese oxides.[45] However, it must kept in mind that artifacts can easily be generated when using the Maxwell equation.[30-42] Before addressing this issue more precisely, let us consider the results of $\Delta S(T)$ obtained from the calorimetric method.

This approach is based on the recording of heat capacity curves, $C(T)$ measured in zero-field and in the external magnetic field H. Assuming that the basic relationship $dS = (CdT)/T$ is always obeyed (i.e, dealing with reversible transformations), the isothermal entropy change is directly computed from the following equation,[46]

$$\Delta S(T,H) = S_H(T) - S_0(T) = \int_0^T \left[\frac{C(T,H)}{T}\right] dT - \int_0^T \left[\frac{C(T,0)}{T}\right] dT \qquad (2)$$

The C(T) curves in H = 0 and H = 20 kOe were measured upon warming after a zero-field cooling down to the lowest reachable temperature, i.e. 2 K in the present case. Below 2K, the C/T values are linearly extrapolated to zero at T = 0 K. The $-\Delta S(T)$ derived using the above process is depicted in Figure 2a, where the main feature is again the presence of a large peak centered at $T_N$. However two main differences emerge when comparing to the magnetic method. First, the height of the peak is substantially higher, reaching a maximum value of 11 J K$^{-1}$ kg$^{-1}$ instead of 7.5 J K$^{-1}$ kg$^{-1}$. Second, the $-\Delta S(T)$ at low-temperatures monotonically reaches zero as the temperature is decreased without exhibiting any crossover to negative values. The first issue is a direct consequence of the limited resolution of the magnetization method. Indeed, the $\Delta S$ value derived from Eq. (1) actually reflects an average over the temperature range $T \pm \delta T$, where $\delta T$ is the spacing between two consecutive $M_T(H)$ curves. In particular, one can check that the maximum value derived from magnetization, $\Delta S (44K) \approx$ -7.5 J kg$^{-1}$ K$^{-1}$, well corresponds to $\left(\int_{44-\delta T}^{44+\delta T} \Delta S_C(T) dT\right)/2\delta T = -7.44$ J K$^{-1}$ kg$^{-1}$, where $\Delta S_C(T)$ is the curve derived from heat capacity. Of course, reducing $\delta T$ minimizes this effect, but this is limited by a growing uncertainty in the derivative of Eq. (1) when the $M_H(T)$ become too close to each other. In reality, the intrinsically higher resolution of the calorimetric method makes it more suitable to derive the peak of $\Delta S(T)$ in case of a sharp transition, as presently found around $T_N$.

In other respects, the absence of positive values in the $\Delta S(T)$ curves derived from $C(T)$ at low-temperatures led us to reconsider the results obtained from the magnetization method. Actually, the question is not about the quality of magnetic data nor the Maxwell equation in itself, but it deals with a possible inappropriate use of it.[6,35,36,46] For instance, it is now widely recognized that such a type of problem is at the origin of a lot of huge $\Delta S$ values reported in the literature, which are in fact artifacts.[30-42] An important point sometimes forgotten is that the Maxwell equation is supposed to hold only in presence of a reversible magnetic behavior.[6,35,36,46] Hence, a first check about the legitimacy of using the Maxwell equation is to compare the M(H) curves recorded either upon field-increasing ("Up" curve) or field-decreasing ("Down" curve). Figure 3 shows some of such magnetic cycles at various temperatures. The M(H) curves recorded at temperatures higher than ~ 30 K (see Figure 3a), i.e. in the temperature range of the transitions, are found to be reversible (or *almost* reversible), authorizing the use of the Maxwell equation. In contrast, at lower temperature (see Figure 3b), a noticeable hysteresis is observed between the field increasing and field decreasing branches of M(H) curves. This behavior becomes much more prominent at low temperatures (below ~ 15 K). Therefore, the Maxwell equation is simply not valid to estimate the $\Delta S$ using Equation (1) at low-temperatures. Looking at Figure 3b, one can notice that the apparent "positive" values of $\Delta S(T)$ reflect a shift of the M(H) "Up" curves towards higher fields as the temperature is decreased. Meanwhile the application of the Maxwell equation on the "Down" series of M(H) curves would just yield vanishing positive values of $\Delta S(T)$ in the same low-temperature regime. In fact, such a low-temperature behavior of the M(H) curves is mainly driven by the pronounced temperature dependence of the coercive field in $Mn_3O_4$,[19] a phenomenon which should not be confused with a genuine MCE.

Finally, a closer look at the $-\Delta S(T)$ curve derived from heat-capacity reveals two additional features which are more clearly evidenced in the logarithmic scale (Figure 2b). Around 40 K (close to $T_1$), one can observe a weak "kink", while a small "dip" appears around 33 K (near to $T_2$). The kink at $T_1$ can be regarded as an additional negative contribution of entropy change superimposed onto the low-temperature side of the main peak around $T_N$. Such a behavior is well corresponding to a shoulder observed around $T_1$ in $-dM/dT$ curve (see inset Figure 1). In terms of heat capacity, the above feature is also consistent with the fact that around $T_1$, $C(T, H=20kOe)$ lies below $C(T, H=0)$ and has its maximum slightly shifted towards higher temperature. At $T_2$, the anomaly looks different, since it rather appears as a "dip" superimposed onto the low-temperature wing of the main peak. It turns out that this feature comes from a peculiar influence of the magnetic field on the $C(T)$ curve around $T_2$. As compared to zero-field, the peak present on the $C(T)$ curve in 20 kOe starts developing at a lower temperature ($\approx 32.5$ K instead of $\approx 33.0$ K) while its maximum is slightly shifted to higher temperature ($\approx 34.3$ K instead of $\approx 34.0$ K). The crossing between the $C(T)$ curves in H=0 and H=20 kOe is close to $\approx 33.6$ K. In terms of MCE, such a field-induced modification in the profile of the transition induces a dip in $-\Delta S(T)$, that is centered at the crossing temperature. Moreover, the height of such a peak is expected to be only a fraction of the entropy jump associated with the first-order transition at $T_2$. By integrating the anomaly on the $C/T(T)$ curves in zero or 2T, this entropy jump is found to be $\sim 0.5$ J $K^{-1}$ $kg^{-1}$, while the height of the dip-like anomaly on the $\Delta S(T)$ curve is indeed smaller, i.e. $\sim 0.1$ J $K^{-1}$ $kg^{-1}$.

With the calorimetric method, one can also obtain the second characteristic quantity of the MCE, i.e. the adiabatic temperature change $\Delta T_{ad}$. These values were derived from the entropy curves using the equation,[46]

$$\Delta T_{ad}(T, 0 \rightarrow H) = \left[ T_H(S) - T_0(S) \right]_{S_0(T)} \qquad (3)$$

The resulting $\Delta T_{ad}(T)$ curve in the case of field change from 0 to 20 kOe is shown on Figure 4. Similarly to the $\Delta S(T)$ curve, the main feature in $\Delta T_{ad}(T)$ is a prominent peak located close to $T_N$. One can also observe a clear signature of the anomaly at $T_2$ in the $\Delta T_{ad}(T)$ plot. It can be noted that the maximum $\Delta T_{ad}$ of 1.9 K for a moderate applied magnetic field of 20 kOe, is large enough to regard $Mn_3O_4$ as a potential refrigerant material. Nevertheless, the width of the $\Delta S(T)$ and $\Delta T_{ad}(T)$ curves are quite small, which would yield only a modest relative cooling power. The best magnetocaloric materials reported so far in the temperature range around 50 K belong to the family of the Laves phases.[3,15] Even though the performances of $Mn_3O_4$ well compare to those of these compounds showing second-order ferromagnetic transitions, slightly better results are observed in the case of first-order transitions. As a matter of fact, the best Laves phase $ErCo_2$ ($T_C = 37$ K) yields a maximum $\Delta T_{ad} = 3$ K for a field change of 20 kOe.[3,15,47,48]

## IV. Conclusions:

In conclusion $Mn_3O_4$ is found to exhibit a substantial MCE around its ferrimagnetic transition at $T_N$ (42.75 K). For a field change of 20 kOe, the peak values of $|\Delta S|$ and $|\Delta T_{ad}|$ are 11 J K$^{-1}$ kg$^{-1}$ and 1.9 K, respectively. The various transitions taking place from $T_N$ down to ~ 33 K are also found to have signatures on the isothermal entropy change. However, these effects are smaller and do not really contribute to widen the operating temperature range for applications in magnetic refrigeration.

This study also well illustrates the advantages of the calorimetric method as compared to the widely used magnetic technique based on the Maxwell equation: First, it is less prone to the generation of spurious features, as those we presently observed in $Mn_3O_4$ owing to the appearance of magnetic irreversibility at low temperatures ; Second, the calorimetric method facilitates a better temperature resolution which allows to detect the small anomalies and avoid underestimating the maximum $\Delta S$ value in case of sharp peaks.

## V. Acknowledgements:

Research at Laboratoire CRISMAT is supported by Indo-French Centre for the Promotion of Advance Research/Centre Franco-Indien pour la Promotion de la Recherche Avancée (IFCPAR/CEFIPRA), the CNRS Energy Program (Research Project "Froid Magnétique") and ENERMAT.

<u>Figure captions</u>

Figure 1. Temperature dependence of the isothermal entropy change derived from the magnetic method, using the field-increasing branches of M(H) for a field variation from 0 to 20 kOe. The inset displays the derivative of the magnetization curve measured in 200 Oe. The arrows show the transitions taking place, from right to left, at $T_N$, $T_1$, and $T_2$.

Figure 2. (a): Main panel: Temperature dependence of the isothermal entropy change for a field variation of 20 kOe, where the "magnetic" ΔS(T) (open squares) is derived from the field increasing branches of M(H), while the "calorimetric" ΔS(T) (filled circles) is derived from C(T) curves recorded upon warming. The inset shows the temperature dependence of C/T in zero-field. (b): Zoomed view of the isothermal entropy change estimated from the heat capacity data, represented in a semi-logarithmic scale. The arrows highlight the anomalies present, from right to left, at $T_1$ and $T_2$.

Figure 3. Isothermal magnetization curves recorded either upon increasing (filled symbols) or decreasing (empty symbols) the magnetic field. (a): High-T regime with curves at 30, 35, 40 and 45 K; (b): Low-T regime with curves at 5, 10, 15 and 20 K.

Figure 4. Temperature dependence of the adiabatic temperature change in $Mn_3O_4$, measured for a magnetic field change of 20 kOe.

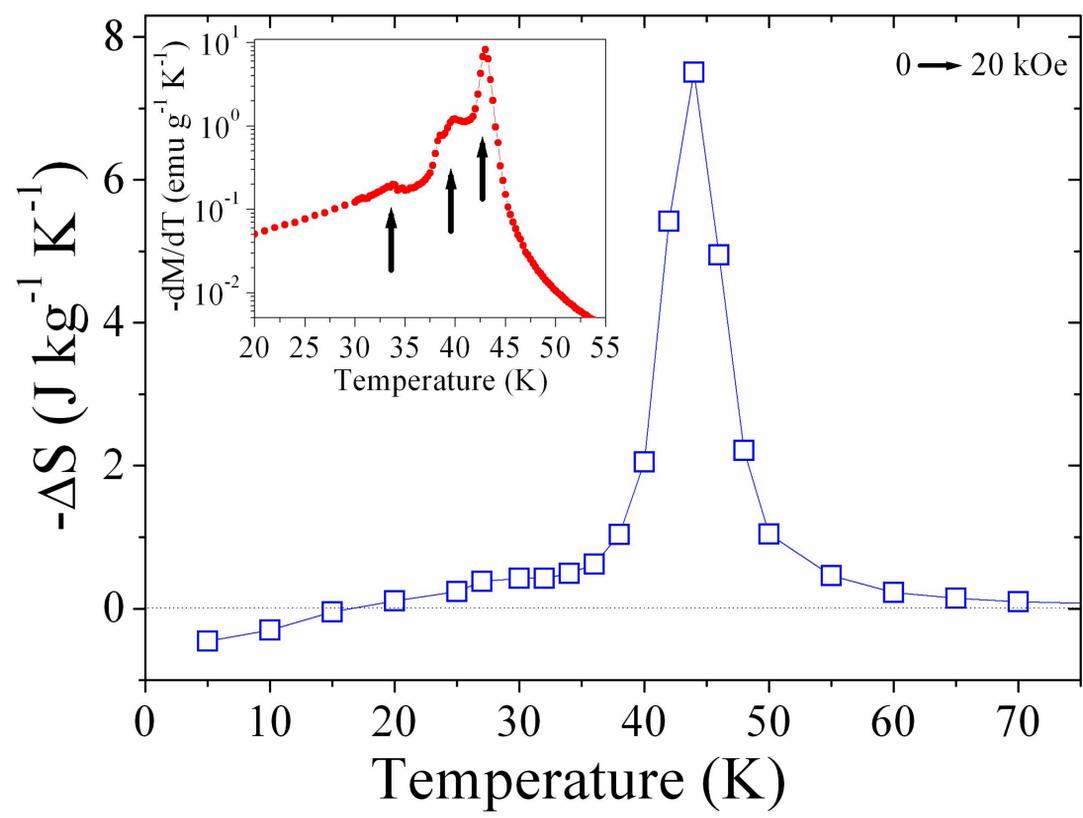

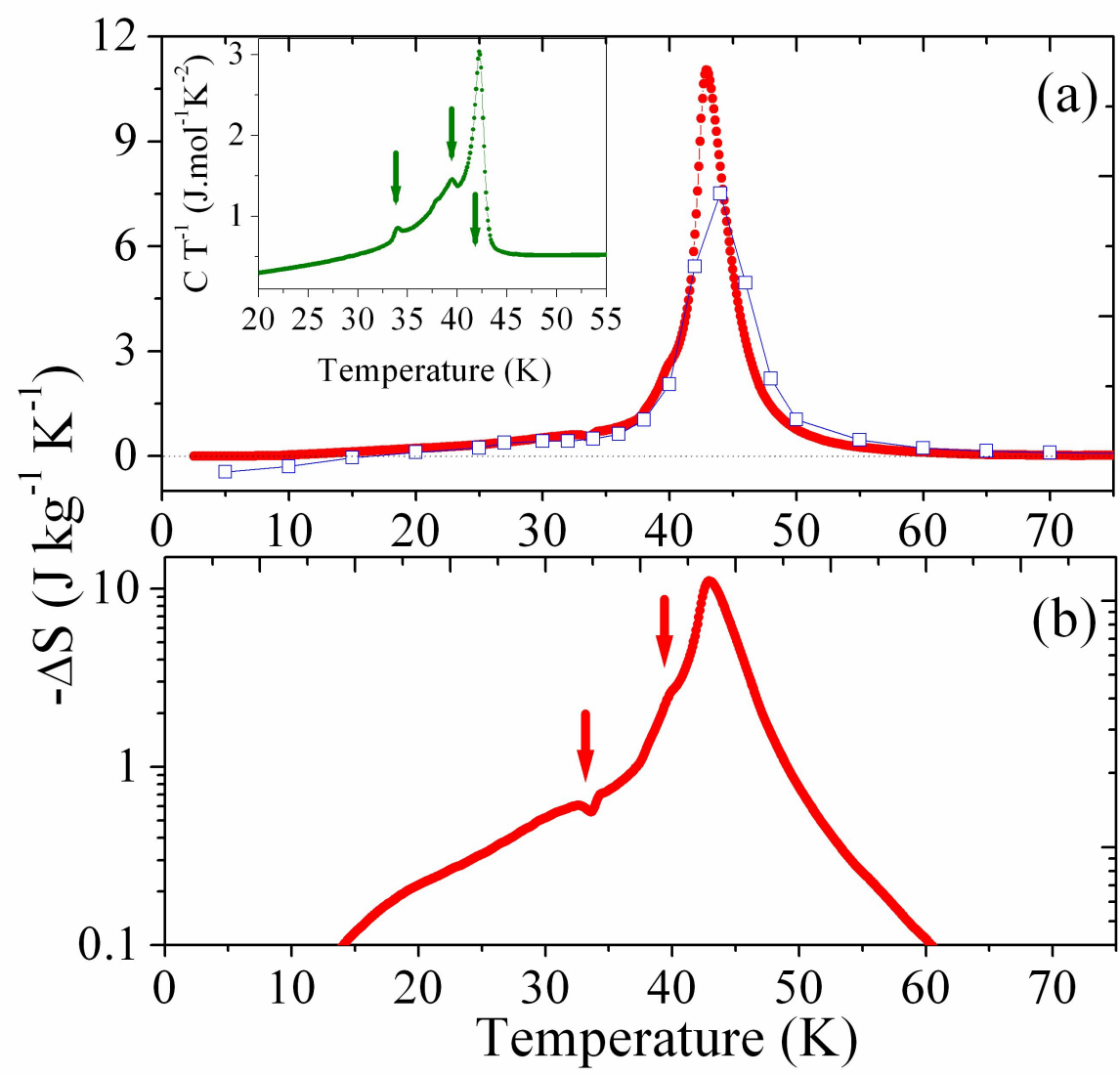

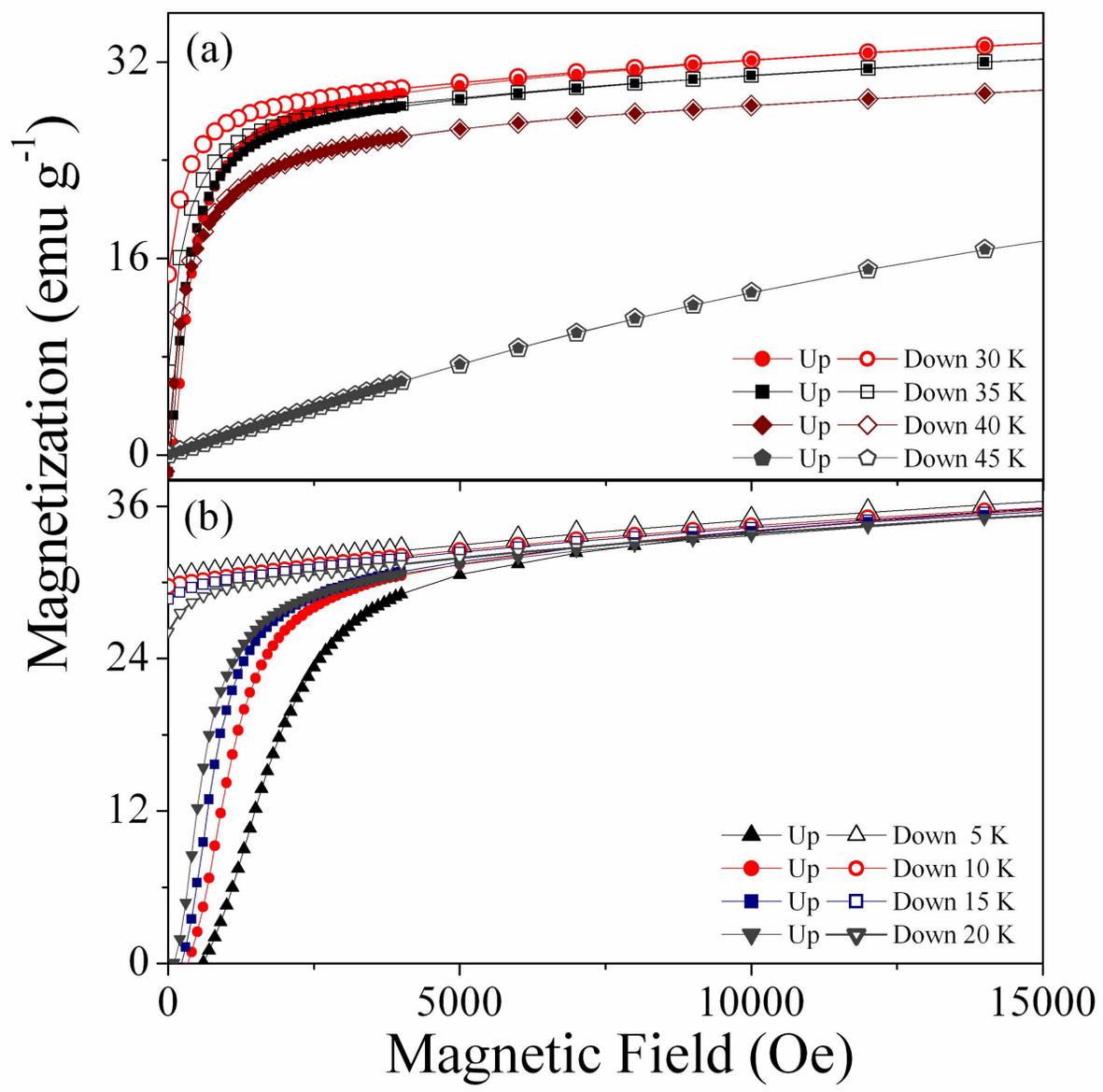

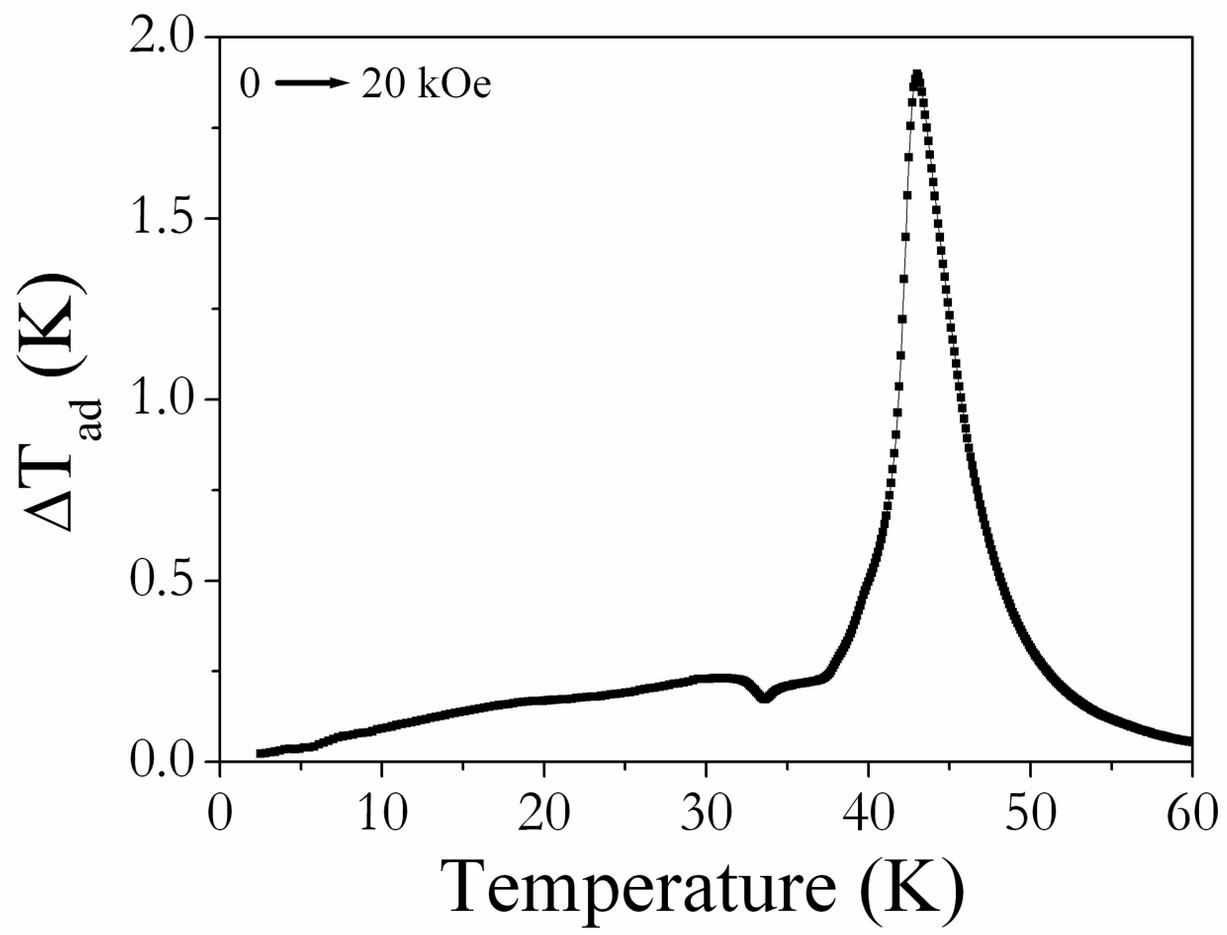